\documentclass[prl,twocolumn,showpacs,preprintnumbers,amsmath,amssymb]{revtex4}

\usepackage{graphicx,color}
\usepackage{multirow}

\begin{document}

\preprint{}

\title{Ferromagnetic proximity effect in F-QDot-S devices}

\author{L.~Hofstetter$^{1,*}$, S.~Csonka$^{1,2,*}$, A.~Geresdi$^{2}$, M.~Aagesen$^{3}$, J.~Nyg{\aa}rd$^{3}$ and C.~Sch\"{o}nenberger$^{1}$}

\affiliation{$^{1}$ Department of Physics, University of Basel, Klingelbergstr. 82, CH-4056 Basel, Switzerland}
\affiliation{$^{2}$ Department of Physics, Budapest University of Technology and Economics, Budafoki u. 6, 1111
Budapest, Hungary} \affiliation{$^{3}$ Niels Bohr Institute, Univ. of Copenhagen, Universitetsparken 5, DK-2100
Copenhagen, Denmark \\
$^{*}$ These authors contributed equally to this work}

\date{\today}

\begin{abstract}
Ferromagnetic proximity effect is studied in InAs nanowire (NW) based quantum dots (QDs) strongly coupled to a
ferromagnetic (F) and a superconducting (S) lead. The influence of the F lead is detected through the splitting of the
spin-1/2 Kondo resonance. We show that the F lead induces a local exchange field on the QD, which has  varying
amplitude and sign depending on the charge states.  The interplay of the F and S correlations generates an exchange
field related subgap feature.
\end{abstract}


\maketitle

Superconductor-ferromagnet (S/F) heterostructures have been investigated extensively in the past and they have
potential impact on the field of spintronics and quantum computation. Due to the different spin ordering of textcolor{red}{a} s-wave
superconductor and a ferromagnet, a large variety of interesting phenomena can be studied at a S/F interface: probing
the spin polarization by Andreev reflection~\cite{Soulen_1998}, $\pi$-junction
behaviour~\cite{Kontos_2001,Zareyan_2001}, ferromagnetically induced triplet superconductivity~\cite{Keizer_2006} or
ferromagnetically assisted Cooper pair splitting~\cite{Beckmann_2004}.

Here we focus on a novel S/F hybrid device where a quantum dot (QD) is incorporated between a S and a F lead.
Recently, the modification of the Andreev reflection of such hybrid sytems has been calculated~\cite{Feng2003,Cao2004}.
Also configurations, where a QD is strongly coupled to two F leads (F-QDot-F) attracted increasing
theoretical~\cite{Martinek2003,Choi2004,Cottet2006a,Cottet2006b} and
experimental~\cite{Sahoo2005,Pasupathy2004,Hamaya2007,Hauptmann2008} attention. It has been
verified~\cite{Pasupathy2004,Hamaya2007,Hauptmann2008}, that in this situation the spin-$\uparrow$ and $\downarrow$
energy levels of the QD can split by an exchange energy, $E_{ex}$ due to the hybridization between the QD states and
the F lead. This proximity ferromagnetism has the same effect on the QD as the presence of an external magnetic field,
$B$. Thus, the exchange splitting is often characterized by the so-called local magnetic exchange field ($B_{ex}$) in
the literature, where $E_{ex} = g\mu_B B_{ex}$. As has been proposed by Martinek~$et \, al.$~\cite{Martinek2005},
$B_{ex}$ can be gate dependent and even change sign. Hence, electric controlled reversal of
the spin occupation of the QD can be achieved. This has very recently been shown in carbon
nanotubes with two F contacts~\cite{Hauptmann2008}.

In this work, we show the influence of F and S correlations on the transport properties of F-QD-S devices. We
demonstrate that already a single F lead introduces a local exchange field which strongly varies for different QD
levels. Furthermore, it even can change its sign within the same charge state. On the other hand, the presence of the S
contact serves as a spectroscopic tool. Concentrating on the subgap transport, a novel mini-gap feature is observed,
which is related to $B_{ex}$.
\\

A typical device is shown in Fig.~1a. We use high-quality molecular beam-epitaxy grown InAs nanowires (NWs). NWs are
dissolved in IPA and deposited on doped Si substrates with $400$~nm insulating SiO$_2$. Afterwards, ohmic contacts at
spacings of $300-500$~nm are fabricated. The F contacts consist of a Ni/Co/Pd trilayer (15~nm/80~nm/10~nm), whereas the
S ones consist of a Ti/Al bilayer (10~nm/110~nm). Prior to metal evaporation, argon gun sputtering is used to remove
the native oxide layer from the nanowires. In agreement with our previous measurements, QDs form between these
contacts, which can be tuned by the voltage applied on the backgate, $V_{BG}$ ~\cite{Csonka2008,Jespersen2006}.
Measurements are performed at a temperature of $25$~mK. A $B$ field is applied parallel to the easy axis of the F
contact which is defined by shape anisotropy (see Fig.~1a). The $B$ field allows to switch the orientation of the F stripe and to
control the size of the superconducting gap. To magnetize the F lead the external field is ramped to $B=+300$~mT and
then back to $B=0$~mT before measuring.

In Fig.~1b the energy diagram of our devices is shown schematically. The F lead induces an asymmetry of the
tunnel coupling ($\Gamma$) of spin-$\uparrow$ and spin-$\downarrow$ electrons to the QD, described by the tunneling
spin polarization $P=(\Gamma_{\uparrow}-\Gamma_{\downarrow})/(\Gamma_{\downarrow}+\Gamma_{\uparrow})$.  This asymmetry
is caused by the difference of the spin-$\uparrow$ and spin-$\downarrow$ electron density of the F lead at the Fermi
energy  and by the tunneling matrix elements of these electrons. E.g. in Ni at $E_F$ two bands couple with opposite
spin-imbalance and very different tunneling matrix elements to the QD~\cite{Mazin1999}. By hybridization, the spin
dependent tunnel coupling generates a spin imbalance on the QD, described as an exchange field. The S
lead is represented by the BCS density of states (DOS) with its energy gap, $\Delta$.

\begin{figure}[bh!]
\centering
\includegraphics[scale=0.4]{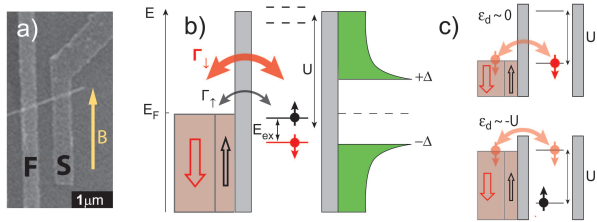}
\caption{(Color online)(a) A SEM picture of a device. The right lead is a Ti/Al bilayer superconductor, while the left
longer one is a Ni/Co/Pd trilayer  ferromagnet. The $B$ field is applied parallel to easy axis of the F lead.  (b) Schematic
view of the F-QDot-S system. The spin degeneracy on the QD is lifted by an exchange splitting induced by the
ferromagnetic proximity effect. (c) Due to the spin polarized charge fluctuations, the spin ground state of the QD is
opposite when the QD occupation fluctuates between 1 and 0 ($\epsilon_d \approx 0$) or 1 and 2 ($\epsilon_d \approx -
U$). } \label{Figure1}
\end{figure}

In order to investigate the ferromagnetic proximity effect, the QD is operated in the strongly coupled regime, which
allows the study of the cotunneling induced manybody spin-1/2 Kondo resonance~\cite{Goldhaber-Gordon1998} at odd
occupation number of the QD. Kondo resonances split into a doublet in a $B$ field according to the Zeeman energy
$E_{\downarrow/\uparrow} = \mp 1/2 g\mu_B B$, thus being also a sensitive tool to visualize $B_{ex}$. Probing the
exchange splitting by the Kondo effect has been demonstrated in C60-molecules~\cite{Pasupathy2004} and carbon nanotubes
based QDs~\cite{Hauptmann2008}. InAs NW QDs have the advantage that the g-factor can be comparable to the bulk value of
g$\approx$15~\cite{Csonka2008}, making them particularly sensitive for studying the level splitting created by
$B_{ex}$. Much smaller external fields are needed to access the regime, where the exchange energy and the Zeeman
splitting are comparable.

The ferromagnetic proximity effect and the therewith connected ground state transition on the QD can be described
in a simple model. Using perturbative scaling analysis for a flatband bandstructure with spin
dependent tunneling rates and including finite Stoner splitting in the leads an analytical formula for the energy
splitting of the spin-$\uparrow$ and the spin-$\downarrow$ is found~\cite{Martinek2005}:
\begin{equation}
eV_{sd} = g\mu_B B + \Delta_0 + (P\Gamma / \pi)ln(|\epsilon_d| / |U + \epsilon_d |).
\end{equation}
Here $P$ is as defined earlier, $\Gamma$ is the coupling to the F lead ($\Gamma=\Gamma_\uparrow+\Gamma_\downarrow$),
$U$ the charging energy of the QD and $\epsilon_d$ the level position of the QD, tunable by $V_{BG}$. $g \mu_B B$ is
the Zeeman splitting due to an external magnetic field, $\Delta_0$ is a Stoner splitting induced shift.  Elaborated NRG
calculations also support the result of Eq.~1~\cite{Martinek2005}. Interestingly, based on Eq.~1, the spin ground state
of the QD is different for $\epsilon_d$ close to $0$ than for $\epsilon_d$ close to $-U$, if $\Delta_0$ or the Zeeman
term are not too big. This change in the ground state of the QD can be explained by the charge fluctuations between the
QD and the F lead (see Fig.~1c). Electrons with majority tunneling spin orientation dominate the charge fluctuations.
Thus, when the QD occupation fluctuates between $1$ and $0$ ($\epsilon_d \approx 0$), the majority tunneling spin
occupies the QD preferably. However, when the occupation fluctuates between 1 and 2 ($\epsilon_d \approx - U$), the
remaining (non fluctuating) spin on the QD has the minority spin orientation of the tunneling
electrons~\cite{Martinek2005,Hauptmann2008}. The transition between these opposite ground states is described by the
sign change of the exchange field in Eq.~1. However, if $\Delta_0$ is the dominant term, a roughly constant splitting
of the Kondo resonance is expected within a charge state.

\begin{figure}[bh!]
\centering
\includegraphics[scale=0.4]{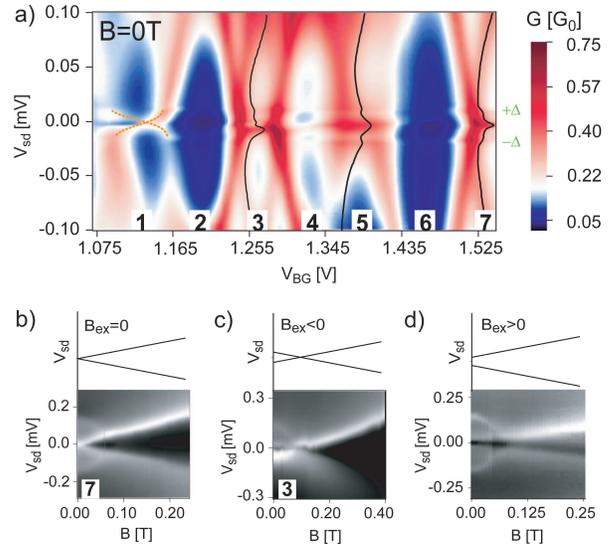}
\caption{(Color online)(a) Differential conductance as a function of $V_{BG}$ and $V_{sd}$ of a F-QD-S device
at $B=0$T. $B_{ex}$ modifies the Kondo
resonances (odd numbered states) differently.
The S lead induces peaks in the conductance at $V_{sd}=\pm \Delta$. $B$ dependence of different charge states: (b)
with no signature of $B_{ex}$ (state 7 in Fig.~2a), (c) with $B_{ex} < 0$, which is compensated by
external $B$ (state 3 in Fig.~2a), (d) with $B_{ex} > 0$, which is enhanced by $B$.  Panel d is measured in
a charge state at $V_{BG}=2.31$~V.}
\label{Figure2}
\end{figure}

In Fig.~2a the differential conductance as a function of $V_{BG}$ and source drain voltage $V_{sd}$ , $G(V_{BG},V_{sd})$,
of several charge states of a studied F-QD-S device is presented. Measurements were performed with standard lock-in
technique with an \emph{ac} excitation of $4 \mu$eV. In accordance with the spin-1/2 Kondo effect, pronounced
conductance is seen around ($V_{sd} = 0$~V) in every odd charge state. However, in these states (labeled with numbers
1,3,5,7) the Kondo resonance shows different signatures due to correlations induced by the F lead. State 7 demonstrates
a spin-1/2 Kondo situation with a single resonance line at $V_{sd}=0$~V. As it is shown in Fig.~2b this zero bias Kondo
resonance splits up linearly with $B$.  In contrast, state 3 exhibits clear signature of F correlations, i.e. the Kondo
resonance has a finite and roughly constant splitting at $B=0$~mT (see black cross-section). Fig.~2c shows, that this
splitting is compensated by $B\approx 64$~mT and split again at higher $B$ fields. Thus $B_{ex}$ has an opposite sign
as the $B$ field. Another type of $B$ dependence of the Kondo ridge is presented in Fig.~2d (measured at
$V_{BG}=2.31$~V), where the zero field splitting of the resonance is further increased by an applied $B$ field. It
means, that $B_{ex}$ is parallel to $B$ for this state. The three markedly different $B$ field behaviors (Fig.~2b-d)
demonstrate that $B_{ex}$ strongly depends on the QD level. Even in a small backgate range the amplitude and the sign
of $B_{ex}$ varies. This observation highlights the particular importance of the coupling of the QD state to the F lead
for the charge fluctuations induced local exchange field~\cite{Martinek2005}. Note, the different $B$ dependencies also
prove that the observed effect can not solely be described by stray field, since the stray field cannot depend on
$V_{BG}$.

The most interesting class among the different exchange field manifestations is the one of state 1 (see Fig.~2a). The
Kondo resonance is also split for this state. However the size of the splitting strongly varies with $V_{BG}$. (The
resonance lines are highlighted with dashed lines in Fig.~2a). Interestingly, the split Kondo resonance lines also
cross the $V_{sd}=0$~mV value inside the charge state. This dependence suggests that within this charge state $B_{ex}$
first gets smaller as $V_{BG}$ is increased. With further increasing $V_{BG}$ it changes its sign and increases further
in the opposite direction. This situation corresponds to the electrically controlled ground state transition on the QD
described by Eq.~1 and Fig.~1c.

\begin{figure}[bh!]
\centering
\includegraphics[scale=0.4]{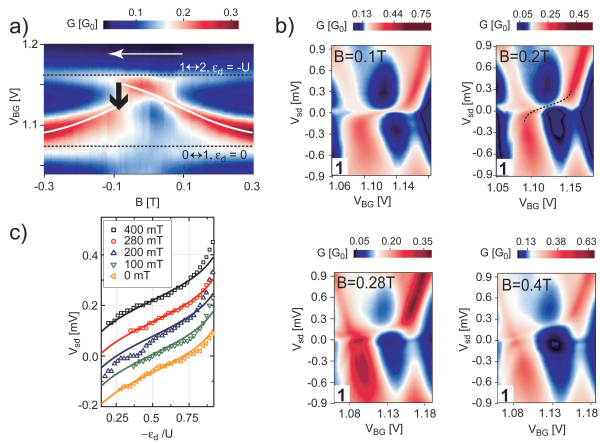}
\caption{(Color online) Measurements on state 1. (a) Differential conductance as a function of $V_{BG}$ and
$B$ measured at $V_{sd}=0$~mV. The two horizontal ridges (dotted lines) are the charge state boundaries. The high conductance ridge is the restored Kondo resonance, where $B=-B_{ex}(V_{BG})$. This
ridge separates the regions where spin-$\uparrow$ or spin-$\downarrow$ is the ground state. The white line is a plot
based on Eq.~1 using the parameters obtained from the fits shown in (c). The white arrow shows the sweeping direction
of $B$. The black arrow points out the position where the magnetization of the F lead changes sign. (b) Colorscale
plots at different magnetic fields. The dashed line highlights the position of the tilted Kondo resonance at $B=0.2$~T.
(c) The evolution of the Kondo resonance is presented for different $B$ fields by  symbols. The lines are fits using
Eq.~1. } \label{Figure3}
\end{figure}

Since $B_{ex}$ in situations like state 1 of Fig.~2a is strongly gate dependent, a measure for the exchange splitting
at different gate voltages can be obtained by measuring the compensation field for which the Kondo peak is restored.
Thus, if one measures $G(V_{BG},B)$  at $V_{sd}=0$~mV, at each $V_{BG}$ signatures of high conductance are expected if
$B=-B_{ex}(V_{BG})$ due to the restored Kondo resonance. Such a plot is presented in  Fig.~3a.  The lower/upper ridge
(see dashed lines) defines the resonance positions when the occupation of the QD level changes between
$0\leftrightarrow 1$ /$1\leftrightarrow 2$. The white arrow indicates the sweep direction of $B$. The high conductance
lines with finite slope show the evolution of the Kondo resonance (marked with a white line). At each $B$ field the
gate position of the restored Kondo resonance defines the border between the spin-$\uparrow$ and the spin-$\downarrow$
ground states. At high external magnetic fields (i.e. $B \geq 0.3$~T) the Zeeman term dominates
 (see Eq.~1).  Therefore, considering that the g-factor of InAs is negative, the spin-$\uparrow$ is the ground state for all
gate values in high field and the Kondo resonance coincides with the border at $\epsilon_d=0$. As the $B$ field
decreases the restored Kondo peak moves towards $\epsilon_d=-U$, opening backgate regions where
spin-$\downarrow$ is the ground state. At $B=0$~mT the spin-$\downarrow$ state dominates, and spin-$\uparrow$ remains only
in a small gate region around $\epsilon_d=-U$  the preferable spin orientation.
The gate voltage value as a function of $B$, where the spin ground state change takes place, can be expressed with
Eq.~1 using the condition of $eV_{sd}=0$. The white line in Fig.~3a shows such a curve, giving a
reasonable agreement with the measurement using the parameters of $|g|=12.3$, $\Delta_0 = -90$~mT and $P\Gamma =
0.22$~meV.

As the polarization of the F lead is switched to the opposite direction by a $B$ field larger than the coercive field
of the contact, $B_{ex}$ also changes its sign.  This appears as a step in the $G(V_{BG},B)$ measurement (see Fig.~3a)
since the splitting turns from $g \mu_B (B+B_{ex})$ to $g \mu_B (B-B_{ex})$. Due to the high g-factor of the InAs QD
state, this is clearly seen in Fig.~3a at the position of the black arrow. Note, that in the vicinity of the
$0\leftrightarrow 1$ border at $B=0$~mT, the ground state is spin-$\downarrow$. At this border the preferable spin
orientation on the QD  is the majority tunneling spin orientation (see Fig.~1c).~\cite{Martinek2005,Hauptmann2008}
 Since the F lead has been previously polarized into spin-$\downarrow$
state by a positive $B$ field, we can conclude, that the polarization of the F lead and the polarization of tunneling
electrons are the same.

Eq.~1 describes the energy difference between the spin-$\uparrow$ and the spin-$\downarrow$ states as a function of
$V_{BG}$. This is equivalent with the energy ($eV_{sd}$), where the Kondo ridge appears. $G(V_{BG},V_{sd})$ of state 1
is also measured in different magnetic fields. As it is seen in Fig.~3b, the position of the more pronounced split
Kondo line moves to higher source drain voltage values as $B$ is increased. The $(V_{BG}$,$V_{sd})$ coordinates of this
line is read out from the measurements, see e.g. the dotted line for $B=0.2$~T.   Fig.~3c summarizes the measured
position of the Kondo ridge by symbols at different $B$ fields. The theory described by Eq.~1 nicely fits the
experiment (see lines) with the parameters used for the plot in Fig.~3a.
\begin{figure}[th!]
\centering
\includegraphics[scale=0.4]{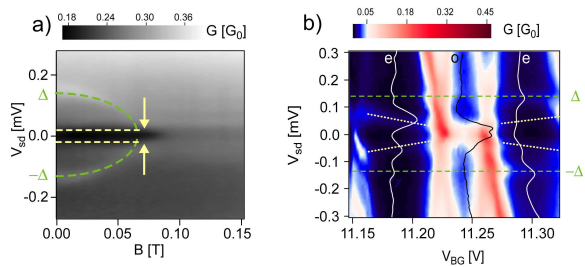}
\caption{(Color online) (a) $G(B,V_{sd})$ measurement: A subgap feature appears (horizontal dashed lines) at the energy
scale of the exchange field. The feature is suppressed above the critical field of the superconductor. (b)
$G(V_{BG},V_{sd})$ at $B=0$~mT showing the relation of the new subgap structure to $B_{ex}$. Inside the superconducting
gap additional resonance lines appear in even charge states (see dotted lines). These resonances change their position
with gate voltage and merge into the exchange split Kondo resonance lines at the border of the odd charge state. The
line graphs show $G(V_{sd})$ cuts in the middle of the three charge states.} \label{Figure4}
\end{figure}

Since our device has a novel hybrid configuration, i.e. a QD connected to one F and one S lead, the interplay of S and
F correlations can be studied. Up to this point, charge states with Kondo physics were investigated, where the Kondo
resonance dominates the low energy behavior also for the S state.  For such states the S lead induces conductance
maxima at $V_{sd}=\pm\Delta$ related to the singularities of the S DOS, which diminish with increasing B field.
However, in several charge states of different devices, where spin-1/2 Kondo resonances are not present (neither in the
S nor in the normal state), an interesting subgap feature appears. This is shown in Fig.~4a, where the $G(B,V_{sd})$
measurement of such a state is presented. The conductance shows in a bias window of $\approx 50~\mu$V significantly
smaller values than in the rest of the superconducting gap (see horizontal dashed line). This mini-gap is clearly
connected to superconductivity since it is suppressed above the critical field of the S lead. On the other hand the
energy scale of the novel subgap  feature coincides with the exchange energy observed in the split Kondo states (see
e.g. Fig.~2d). It suggests that the subgap is also exchange field related. Further evidence for such an relation is
shown in the $G(V_{BG},V_{sd})$ measurement at $B=0$~mT on an other device (see Fig.~4b). The middle charge state with
odd electron filling ($o$) shows a slightly split Kondo resonance, while the two even states ($e$) demonstrate the new
subgap feature. As it is shown by the dotted lines, the mini-gap lines of the even charge states merge into the split
Kondo resonance at the border between the even and the odd states. Thus, superconductivity induced transport processes
in charge states with an even electron number are presented, which seem to be correlated with the exchange splitting.

Concluding, our measurements demonstrate that ferromagnetic proximity effect is indeed present in InAs NW based F-QD-S
devices. A single F lead induces a local exchange field on the QD, which is strongly level dependent: it even changes
sign for a single charge state. Demonstration of controlling the spin ground state of the QD by electric means makes
the F-InAs QD-S system a promising building block for spin correlation studies, if implemented into a Cooper pair
splitter device~\cite{Hofstetter2009,Herrmann2010}.

We thank L. Borda, V. Koerting for fruitful discussions and C.B. Soerensen, III-V Nanolab, Niels Bohr Institute for MBE
growth. This work has been supported by the Swiss NSF, the NCCR on Nanoscale Science, the Danish Natural Science
Research Council, the OTKA-Norwegian Financial Mechanism NNF 78842, OTKA 72916 and the EU M.C. 41139 projects. S. C. is
a grantee of the Bolyai Janos Scholarship.

\end{document}